\documentclass[conference]{IEEEtran}
\IEEEoverridecommandlockouts
\usepackage{cite}
\usepackage{amsmath,amssymb,amsfonts}
\usepackage{algorithmic}
\usepackage{graphicx}
\usepackage{textcomp}
\usepackage{xcolor}
\usepackage{xurl}
\usepackage{bm}
\ifCLASSOPTIONcompsoc
  \usepackage[caption=false,font=normalsize,labelfont=sf,textfont=sf]{subfig}
\else
  \usepackage[caption=false,font=footnotesize]{subfig}
\fi
\def\BibTeX{{\rm B\kern-.05em{\sc i\kern-.025em b}\kern-.08em
    T\kern-.1667em\lower.7ex\hbox{E}\kern-.125emX}}
\begin{document}

\title{Using a Drone Sounder to Measure Channels for Cell-Free Massive MIMO Systems

\thanks{This material is supported by KDDI Research, Inc. and the National Science Foundation (ECCS-1731694 and ECCS-1923601). The full address of ESPOL Polytechnic University is Escuela Superior Politécnica del Litoral, ESPOL, Facultad de Ingeniería en Electricidad y Computación, Km 30.5 vía Perimetral, P. O. Box 09-01-5863, Guayaquil, Ecuador. J. Gomez-Ponce is partially supported by the Foreign Fulbright Ecuador SENESCYT Program.}
}

\author{\IEEEauthorblockN{Thomas Choi\IEEEauthorrefmark{1}, Jorge Gomez-Ponce\IEEEauthorrefmark{1}\IEEEauthorrefmark{2}, Colton Bullard\IEEEauthorrefmark{1}, Issei Kanno\IEEEauthorrefmark{3}, Masaaki Ito\IEEEauthorrefmark{3},\\Takeo Ohseki\IEEEauthorrefmark{3}, Kosuke Yamazaki\IEEEauthorrefmark{3}, and Andreas F. Molisch\IEEEauthorrefmark{1}}
\IEEEauthorblockA{\IEEEauthorrefmark{1}University of Southern California,
Los Angeles, United States} 
\IEEEauthorblockA{\IEEEauthorrefmark{2} ESPOL Polytechnic University, Guayaquil, Ecuador} 
\IEEEauthorblockA{\IEEEauthorrefmark{3}KDDI Research, Inc., Saitama, Japan} 
%\and
}

\maketitle

\begin{abstract}
Measurements of the propagation channels in real-world environments form the basis of all realistic system performance evaluations, as foundation of statistical channel models or to verify ray tracing. This is also true for the analysis of cell-free massive multi-input multi-output (CF-mMIMO) systems. However, such experimental data are difficult to obtain, due to the complexity and expense of deploying tens or hundreds of channel sounder nodes across the wide area a CF-mMIMO system is expected to cover, especially when different configurations and number of antennas are to be explored. In this paper, we provide a novel method to obtain channel data for CF-mMIMO systems using a channel sounder based on a drone, also known as a small unmanned aerial vehicle (UAV). Such a method is efficient, flexible, simple, and low-cost, capturing channel data from thousands of different access point (AP) locations within minutes. In addition, we provide sample 3.5 GHz measurement results analyzing deployment strategies for APs and make the data open source, so they may be used for various other studies. To our knowledge, our data are the first large-scale, real-world CF-mMIMO channel data.
\end{abstract}


\section{Introduction}
\subsection{Motivation} \label{Motivation}
In contrast to the traditional \emph{cellular} system where the signal-to-interference-plus-noise ratio (SINR) varies significantly depending on where the user equipments (UEs) are located within a cell, the \emph{cell-free massive multi-input multi-output} (CF-mMIMO) system can provide an almost uniform quality of services to all UEs by abolishing cell boundaries and distributing many base station (BS) antennas across a wide area in forms of access points (APs) \cite{ngo2017cell-free, interdonato2018ubiquitous, demir2020foundations}. While many studies analyzed how to scale, optimize, and deploy the CF-mMIMO system in a pragmatic manner, the propagation channels used in the analyses were either a) synthetic channels based on statistical channel models (uncorrelated/correlated Rayleigh/Rician) or b) simulated data based on behaviors of electromagnetic waves within a selected environment (ray tracing), whose accuracy, in particular with respect to angular dispersion, is uncertain. When actual real-world measurements were used, the number of APs was small (see Section \ref{Related Works}).

Nevertheless, the measured channel data between \emph{tens to hundreds} of APs and multiple UEs distributed across a \emph{wide area} are needed to accurately model real-world channels for CF-mMIMO systems. This paper presents a novel approach to such large-scale channel measurements that drastically reduces efforts and costs while escalating the data volume: using a drone to create a fast-moving virtual array.

\subsection{Related Works} \label{Related Works}
There are several experimental works that study cooperation of multiple BS antennas distributed across a wide area.\footnote{Relevant studies might use the framework of CF-mMIMO, or might employ other names, such as distributed MIMO, network MIMO, cooperative multi-point (CoMP), distributed antenna system (DAS), etc.} One way to measure propagation channels in such systems is to use a ``full'' real-time system with each antenna having an individual radio-frequency (RF) chain and transceiver, while possibly the APs at different locations are connected through optical fiber back-haul \cite{hiltunen20175g, shepard2018argosnet}. While such measurement system would be the closest to the actual deployed system, its main weaknesses are a) difficulty in installing the back-haul network, b) complexity in terms of calibration, synchronization, and operation, and c) expense scaling with the large number of antennas, and thus transceivers, considered in the CF-mMIMO systems, quickly becoming prohibitive.

An alternative method is to use a \emph{single} RF chain and RF over fiber modules connected to a RF switch \cite{gordon2014experimental, sezgin2019evaluation}. This ``switched'' real-time system allows easier calibration and synchronization while many APs are distributed across a wide area using the low-loss fiber cables. Yet, its price still scales with the number of APs and it is still difficult to make measurements in multiple settings due to challenges of installing and managing many cables and antennas.

The last method is to use a virtual array \cite{alatossava2008measurement, fernandez2008comparison, jungnickel2009capacity, sheng2011downlink,choi2020co-located}, where one antenna (or a co-located antenna array) is used as an AP and another antenna is used as a UE. Such a system operates by fixing the location of the UE antenna and moving the AP antenna to selected AP locations. The UE antenna then moves to the next location and the process repeats. This method is popular, especially in small settings, due to its simplicity and low-cost. However, a key bottleneck is the effort in carrying and installing the transceiver emulating the AP to many different locations. Indeed, all previous measurement-based studies using these three methods involved only a small number of APs. Hence, an efficient, flexible, simple, and low-cost method to measure real-world channels for CF-mMIMO systems is necessary.

\subsection{Contributions}
We describe a new channel measurement method for the CF-mMIMO systems based on a channel sounder flying on a drone, creating a distributed virtual array that can measure channels from thousands of AP locations in a few minutes.\footnote{We stress that we did not build a CF-mMIMO system prototype, but that we obtained real-world CF-mMIMO channel data using the sounder. The measured channels are independent of the sounder hardware, as the responses of the sounder are compensated through careful sounder calibration. The channels do not have to be measured in real-time, if the channels stay constant during the measurement.} To our knowledge, this is the first channel measurement dedicated to CF-mMIMO analysis with such many possible AP locations, and the first time a drone is used to measure such channels.\footnote{While several papers, e.g., \cite{dhekne2017extending}, explored the possibility of using drones as aerial APs, there are no investigations using drones to measure channels for the CF-mMIMO system.} This sounding methodology a) quickly captures channel data from many APs to a UE, b) accesses any outdoor AP location, c) is easy to calibrate and operate using only a single RF chain on each AP/UE end, and d) costs little in terms of both labor and equipment. A sample 3.5 GHz channel measurement campaign is conducted in an outdoor setting at the University of Southern California (USC) campus, and measurement results with insights to realizing a CF-mMIMO system are given. 

\subsection{Reproducible Research}
We encourage researchers to use the real-world measurement data for various CF-mMIMO analyses by making the data open source. The data and the simulation results of this paper are available at: \url{https://github.com/tomathchoi/drone_CF-mMIMO}.

\section{Channel Sounding Methodology}
The channel sounder we use consists of 1) a transmitter (TX) on a drone with a lightweight software-defined radio (SDR) and a single dipole antenna and 2) a receiver (RX) on the ground with a cylindrical antenna array, digitizer, and storage, which are heavier and bulkier than the TX.\footnote{The operating frequency is 3.5 GHz, bandwidth is 46 MHz, length of the sounding sequence is 2301, and output power is 28dBm.} We assume the aerial TX serves as an AP and the ground RX serves as a UE\footnote{In our specific sounder, the RX consists of a dual-polarized cylindrical antenna array with 128 ports \cite{ponce2021air}. Since we assume single-antenna UEs, by selecting one port at a time, we can have 128 UE realizations per RX position. Therefore, using one antenna as a UE will not change the methodology given in this paper.}%Instead of treating each port as a separate realization, the ports can also be used together for various other evaluation purposes, such as direction-of-arrival detection, polarization analysis, multi-antenna UE study, small-scale fading statistics, etc., which are not covered in this paper.}
because a) APs are usually placed higher than UEs and b) the number of APs is assumed to be larger than the number of UEs (the TX sounder on a drone can move to many locations faster than the RX sounder on the ground).

\begin{figure}[t!]
     \centering
     \includegraphics[width=0.8\linewidth]{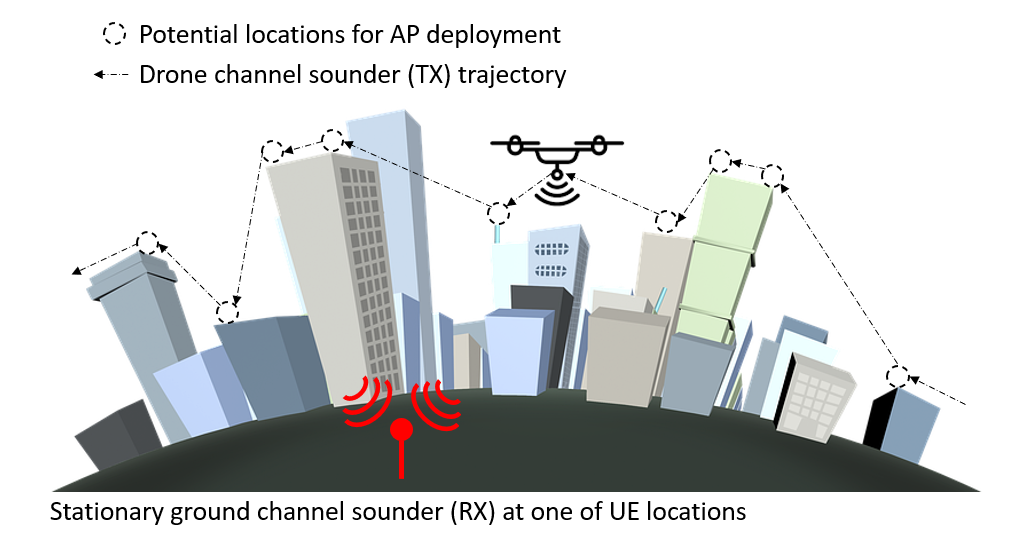}
     \caption{Proposed channel measurement method for a CF-mMIMO system}
     \label{fig:drone_method}
\end{figure}

Measurements proceed as follows: in a selected environment, we first decide at which locations to place the APs and UEs. We fix the RX at the location of the first UE and fly the TX along a trajectory that includes locations of all APs, as shown on Fig. \ref{fig:drone_method}. The measured channels from a single trajectory thus contain the channels between all APs and a single UE. Because the drone flight can be repeated in an automated fashion, the TX can fly the same trajectory repeatedly. We can thus move the RX to the location of a different UE and repeat the same trajectory to obtain the channel data between the same APs and the different UE for a multi-user CF-mMIMO system analysis (the reproducibility of the trajectory will be discussed in Section \ref{channel gains}).

This measurement method has following advantages:
\begin{enumerate}
    \item \emph{Boundless AP locations}: The drone, with its small body, can reach any position at any height conveniently and quickly by using a simple mobile application, as long as the Federal Aviation Administration (FAA) safety rules are followed.\footnote{While a similar method can be created using an automobile with an extendable mast, such a vehicle is more expensive. Furthermore, it can only reach AP locations close to a driving/parking lane in a street.} Such capability is especially useful when measurements for several different CF-mMIMO systems are conducted in multiple environments located far apart, since cumbersome installations of AP antennas at multiple rooftops and masts are not necessary.

    \item \emph{Fast measurement speed and a large dataset}: Although this method is technically the same as the virtual array method using one AP antenna and one UE antenna as described in Section \ref{Related Works}, thousands of AP locations can be swept within several minutes on a drone. In our measurement, the RX captures channel data every 50ms and the TX moves at 4m/s. With such measurement speed, channel data from 1200 AP locations distributed across 240m range to a UE can be measured per minute. We can either sample some of the spatial points among the whole drone trajectory to place a selected number of APs for a considered CF-mMIMO system, or utilize the ample size of the dataset to conduct data-hungry statistical analyses or machine learning applications.
    
    \item \emph{Easy to operate and affordable}: Setting up and operating the \emph{full} or the \emph{switched} sounder mentioned in Section \ref{Related Works} with many AP/UE antennas positioned at multiple different locations simultaneously is complex, and prone to hardware failures if not properly managed. Additionally, the cost of the RF equipment including the transceivers, clocks, cables, switches, amplifiers, filters, and antennas is significant, scaling linearly with the number of APs/UEs. In contrast, measuring channels using a drone sounder is simple and cost-efficient, requiring only a single RF chain which does not require any synchronization due to the virtual array principle.
\end{enumerate}

There are, however, some assumptions and limitations of this measurement method which also must be addressed:

\begin{enumerate}
    \item \emph{Channel coherence in a fast fading channel}: While all APs serve multiple UEs simultaneously in an actual CF-mMIMO system, we measure channels between many APs and a single UE over several minutes of flight time. Furthermore, measurements for different UEs may span different days. Because the channels for APs/UEs can lie in different coherence blocks, dynamic channel conditions such as trucks blocking line-of-sight (LOS) path cannot be accounted for. To minimize such effects, the measurements should be performed at times where the number of such mobile blockers/scatterers is minimal.

    \item \emph{Using channels of multiple UEs together}: We conduct multiple flights of the drone on the same route to measure channels for different UE positions. The drone will not be at the same location during each flight (error on order of wavelength or more), so analysis that requires phase coherence between different UE locations is challenging. This can be overcome by expanding the RX to operate multiple UE antennas simultaneously \cite{choi2021uplink}, which is easier than operating a larger number of AP antennas simultaneously, since RXs are on the ground. 
    
    \item \emph{Drone limitations and effects}: Because the drone has limited power and weight capacity, the TX payload must not be power-hungry or heavy. Performance measures like the bandwidth, output power, number of antennas, and clock accuracy are traded off for lighter hardware with sub-optimal performances. Also, the drone body and the vibration coming from drone hovering may alter the antenna gain and pattern \cite{badi2019experimental}.\footnote{The measurement accuracy of our drone sounder is discussed in \cite{ponce2021air}.}
\end{enumerate}

\section{Measurement Campaign} \label{campaign}
\begin{figure}[!t]
    \centering
    \subfloat[UE1]{\includegraphics[height=6.8cm]{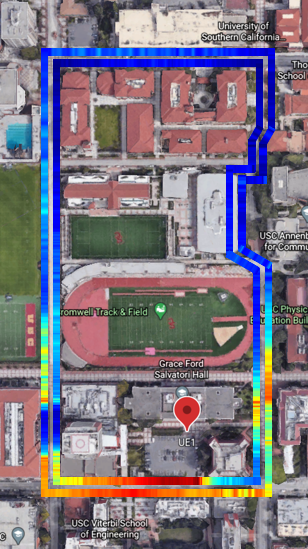}%
    \label{fig:measurement1}}
    \subfloat[UE2]{\includegraphics[height=6.8cm]{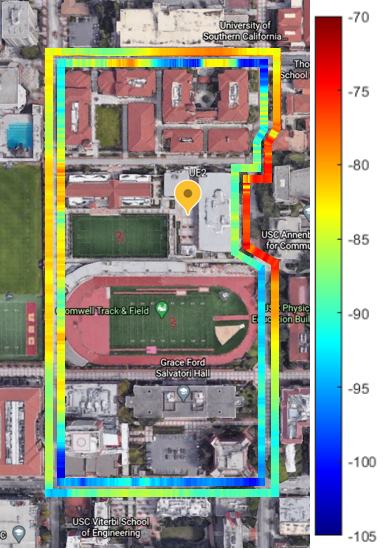}%
    \label{fig:measurement2}}
    \\
    \subfloat[UE3]{\includegraphics[height=6.8cm]{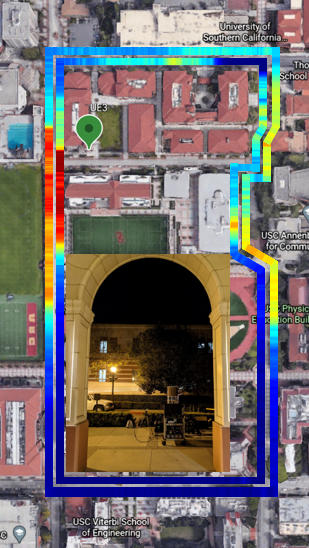}%
    \label{fig:measurement3}}
    \subfloat[UE4]{\includegraphics[height=6.8cm]{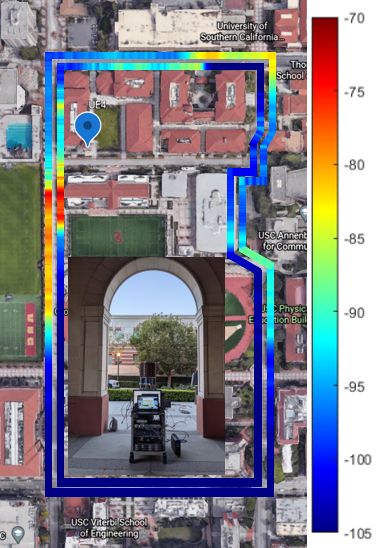}%
    \label{fig:measurement4}}
    
    \caption{Channel measurement environment at USC: colorbars show channel gains (in dB) between all AP locations (drone positions) along the trajectory at 35m (inner loop) / 70m height (outer loop) and four different UE locations.}
    \label{meas_env}
\end{figure}

Using our drone-based sounder that was outlined above and described in more detail in \cite{ponce2021air}, we conducted a CF-mMIMO measurement campaign at USC University Park Campus (Fig. \ref{meas_env}).\footnote{The northeast corner of the drone trajectory on Fig. \ref{meas_env} is bent to avoid hitting tall trees in the area.} The selected measurement area has a dimension of about 400m by 200m. The TX (AP) was flown on a loop trajectory at 35m (rooftop AP) and 70m (aerial AP \cite{dhekne2017extending}) heights. After the trajectory measurements at two heights at a single UE location, the RX was moved to a new UE location and the TX repeated the same two trajectories. The measurement was conducted during dawn time to minimize the number of cars and people, and was conducted over three different days (UE1 on the first day, UE2 on the second day, and UE3/UE4 on the third day). The channel transfer function was captured every 50ms, and the drone moved with 4m/s (measurement every 20cm) for little more than 5 minutes, resulting in over 6000 transfer functions covering more than 1200m of AP locations per drone height per UE.

UE1 was placed at a parking lot near the southeast corner of the area surrounded by tall buildings at west, north, and east (Fig. \ref{fig:measurement1}), UE2 was near the center of the area away from tall buildings for most parts except for a building on the east (Fig. \ref{fig:measurement2}), and UE3/UE4 were at the northwest corner of the area near the road (Fig. \ref{fig:measurement3} and Fig. \ref{fig:measurement4}). UE3 and UE4 were placed close to one another to observe multi-user performance when the UEs are spaced close together, as well as differences in channel characteristic when the UE is placed outside a building versus under a protruding roof. 

While actual CF-mMIMO systems are likely to have a much larger number of UEs, this initial study focuses on analyzing the simple scenario with a small number of UEs in order to straightforwardly study the feasibility of the unique channel sounding method for various CF-mMIMO system analyses. Extensive measurement campaigns at larger scales with more UEs are presented in \cite{choi2021uplink}.

\section{Cell-Free Massive MIMO System Analysis}
\subsection{Channel Gains for Each UE} \label{channel gains}
Channel gain is an important parameter as it describes the quality of the channel between an AP and a UE, and determines other parameters such as signal-to-noise ratio (SNR), SINR, spectral efficiency, etc. We define channel gain between AP $l$ and UE $k$ at $i$th realization as $|h_{l,k,i}|^2$.\footnote{We treat each frequency point from the measurement as a different realization.} Fig. \ref{meas_env} plots channel gains\footnote{On Fig. \ref{meas_env}, omni-directional antenna pattern was synthesized for each UE using multiple vertically polarized ports of the RX antenna array to account for multipath components from all directions.} between the AP locations across the trajectories at two heights and four UE locations, averaged over all realizations.

For UE locations that are surrounded by many buildings (UE1, UE3, and UE4), channel gain is higher (in red) if there is a LOS path toward an AP location, but lower (in blue) if there is shadowing from the buildings or if the AP location is far from the UE location. UE2, which has LOS paths to many areas due to being away from tall buildings, generally shows higher channel gains across the trajectories than other UEs. Because having a LOS channel to multiple APs in metropolitan areas with high buildings is difficult, it is best to spread the APs across many locations, in order for the UE to have high channel gain to at least one UE, as suggested by the CF-mMIMO idea \cite{interdonato2018ubiquitous}. In contrast, fewer APs or even one BS may be good enough in rural areas without tall buildings.

Heights of the APs must also be considered when deploying APs for CF-mMIMO systems. If we compare the 35m measurement and the 70m measurement, the 70m trajectories generally have more regions with high channel gains (see, e.g., the east side of UE2 and the northeast side of UE4) because shadowing between APs and UEs surrounded by buildings can sometimes be eliminated by simply increasing the height of the AP. While the trajectories we observe are both beyond 30m height, flying at lower heights is expected to reduce the regions with high channel gains even further. This qualitatively suggests that the required density of the APs during the deployment will be strongly correlated with the heights of the APs.

One important aspect of our sounding method is the reproducibility of the measurement. To observe this, we repeated the same trajectory measurement twice for UE2 over two different days, and compared the channel gains over the same trajectories in a following way:

\begin {equation}
    G_{l,2,t}[\mathrm{dB}] = 10\mathrm{log}_{10}(\frac{\sum_{i=1}^{F}|h_{l,2,i,t}|^2}{F})
\end{equation}

\begin{equation}
    \mathrm{err}_{\mathrm{RMS}} = \sqrt{\frac{\sum_{l=1}^{L}(G_{l,2,1}[\mathrm{dB}]-G_{l,2,2}[\mathrm{dB}])^2}{L}}
\end{equation}
where $h_{l,2,i,t}$ is a channel between AP $l$ and UE $2$ at realization $i$ for trial $t$, $G_{l,2,t}[\mathrm{dB}]$ is the sum of channel gains between AP $l$ and UE $2$ averaged over all realizaitons for trial $t$ expressed in dB scale, and $\mathrm{err_{RMS}}$ is the RMS error between two $G_{l,2,t}[\mathrm{dB}]$s over the whole trajectory containing $L$ spatial points. 

The resulting RMS error was about 2.63dB. While this variation is not too big, the comparison of complex channel resulted in much larger variation, which suggests that while the magnitude does not vary as much, the phase will vary when we repeat the same trajectory. Thus, if the exact phase relationship between different UEs is required, simultaneous measurement with multiple UEs is preferable. However, if the UE locations are widely separated and their phase relationships are essentially random, the details are not relevant; this is also the case for the analysis in Sec. IV.C. 

\subsection{Single-User Uplink SNR} \label{uplink_SNR}
We now look at uplink SNR in a single-user case where only one UE is served by multiple APs at a given channel resource (time/frequency slot). Among all AP positions, we select a given number of APs,\footnote{The selection of APs can be random or deterministic.} and combine the SNR between the UE %\footnote{For analysis in Section \ref{uplink_SNR} and \ref{SINR}, we randomly select one antenna port, accounting for random orientation of the UE patch antenna.} 
and all selected APs through maximal-ratio (MR) combining to get the total SNR. We consider the measured channel as the ground-truth channel in our evaluations. While the channel measurement is conducted from the AP (TX) side to the UE side (RX), we can still evaluate the \emph{uplink} SNR because the channels are reciprocal and independent of whether the operation is uplink or downlink as long as the responses of RF chains measured through the back-to-back calibration is eliminated from the total channel responses of the channel measurement system.

We assume the transmit power from each UE ($p$) is 0dBm and the noise power at each AP during the uplink ($\sigma_{\mathrm{ul}}^2$) is -90dBm. Following \cite{demir2020foundations}, if there are $L$ single-antenna APs which are designated for UE $k$, the resulting uplink SNR for UE $k$ in a cell-free system is:

\begin{equation}
    \mathrm{SNR}_k = \frac{p}{\sigma_{\mathrm{ul}}^2}\sum_{l=1}^{L}\sum_{i=1}^{F}\frac{|h_{l,k,i}|^2}{F}.
\end{equation}

\begin{figure}[!t]
    \centering
    \subfloat[UE1]{\includegraphics[width=0.8\linewidth]{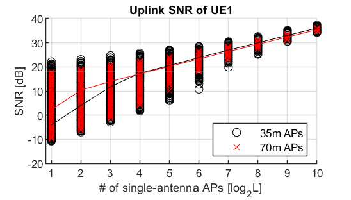}%
    \label{fig:snr_dist1}}
    \\
    \subfloat[UE2]{\includegraphics[width=0.8\linewidth]{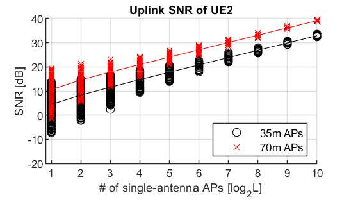}%
    \label{fig:snr_dist2}}
    
    \caption{Distribution of UE1/2 uplink SNR for cell-free case per number of APs at two different AP heights}
    \label{fig:snr_dist}
\end{figure}

Fig. \ref{fig:snr_dist} plots SNR versus number of single-antenna APs, at two different AP heights for UE1 and UE2, based on the measured channel data. The number of single-antenna APs varies by power of 2, going from 2 all the way up to 1024. The APs are picked at random among all AP locations, with 10000 different AP combinations chosen from the set of measured locations. We again use synthesized omnidirectional antenna on the UE side, where channel gains are averaged over multiple realizations. The first observation is that as the number of APs increase, the median SNR (shown by the solid lines) increases and the variation of the SNR values decreases (for example, standard deviation reduces from when 8.5dB to 0.4dB for UE1/AP35m) for all considered cases. This is because as we increase the number of APs, the APs would likely cover most parts of the measurement setting, resulting in similar performance even when we are picking the APs at random. We also see that the variance is smaller when the APs are at 70m height than 35m height. This is because the channel gain varies less along the trajectory at 70m height than 35m height (see Fig. \ref{meas_env}). 

Another observation is that UE2 generally has better performance than UE1. While UE1 only has limited regions with high channel gains, UE2 contains many more regions with high channel gains. Hence, it is likely that the selected APs will contain at least one path with high channel gain to UE2, while it is not the case for UE1. UE2 also has less variance in SNR than UE1, since the channel gains are more uniformly high in contrast to UE1 where the variance of channel gains along the trajectory is very high. SNR values of UE3 and UE4, while omitted, provide similar behaviors in variations as UE1, since the distribution of channel gains is similar as shown in Fig. \ref{meas_env}.

%The next observation we make is in regards to the deployment strategies for APs. Fig. \ref{fig:pdf_SNR} shows probability distribution of SNR values for UE1 when served by 64 APs at 35m height with the APs distributed randomly versus evenly, meaning we divide the whole trajectory to 64 equally spaced subset trajectories and pick one AP at random per subset trajectory. We notice that the mean SNR increases and the SNR variation is much less when the APs are distributed evenly. A UE can experience very bad or very good channels if all APs are concentrated, such as in randomly distributed case. Since the goal of CF-mMIMO is to provide uniform quality of service, distributing APs evenly can help achieve such goal better.

%\begin{figure}[b!]
%     \centering
%     \includegraphics[width=1\linewidth]{figures/Fig_4.png}
%     \caption{Distribution of UE1 uplink SNR for cell-free case using two different AP deployment strategies}
%     \label{fig:pdf_SNR}
%\end{figure} 

\subsection{Multi-User Uplink SINR} \label{SINR}
Now we look at the case when multiple UEs are served together by all APs within the same time/frequency slot. This time, we use a randomly selected vertically polarized port from the cylindrical array as a UE instead of synthesizing an omni-directional antenna, meaning UE antenna has a directivity and is pointing at random direction. The parameter we look at is the SINR, as the interference from other UEs may reduce the achievable spectral efficiency. Again, following \cite{demir2020foundations}, the SINR is computed as:

\begin{equation}
    \begin{split}
    \mathrm{SINR}_k &= \frac{|\bm{\mathrm{v}}_k^\mathrm{H}\bm{\mathrm{h}}_k|^2p}{\bm{\mathrm{v}}_k^{\mathrm{H}}(p\sum\limits_{\substack{i=1\\i\neq k}}^{K}\bm{\mathrm{h}}_i\bm{\mathrm{h}}_i^{\mathrm{H}}+\sigma_{\mathrm{ul}}^2\bm{\mathrm{I}}_M)\bm{\mathrm{v}}_k} \\ & \leq p\bm{\mathrm{h}}_k^{\mathrm{H}}(p\sum\limits_{\substack{i=1\\i\neq k}}^{K}\bm{\mathrm{h}}_i\bm{\mathrm{h}}_i^{\mathrm{H}}+\sigma_{\mathrm{ul}}^2\bm{\mathrm{I}}_M)^{-1}\bm{\mathrm{h}}_k
    \end{split}
\end{equation}

\noindent where $M$ is the number of BS antennas (in our analysis, also the number of single-antenna APs), $\bm{\mathrm{v}}_k$ is the $M\times1$ combining vector for UE $k$, $\bm{\mathrm{h}}_k$ is the $M\times1$ channel vector between UE $k$ and all APs, $(\cdot)^{\mathrm{H}}$ is the Hermitian operation, and $\bm{\mathrm{I}}_M$ is the $M\times M$ identity matrix. The upper bound is achieved by computing $\bm{\mathrm{v}}_k$ via a generalized Rayleigh quotient \cite{bjornson2017massive}, which we call optimum combining:\footnote{It is very similar to the minimum mean square error (MMSE) combining vector, but without the channel estimation error correlation matrix in the denominator as we do not consider channel estimation.}

\begin{equation} \label{SNR}
    \bm{\mathrm{v}}_k = (p\sum\limits_{\substack{i=1\\i\neq k}}^{K}\bm{\mathrm{h}}_i\bm{\mathrm{h}}_i^{\mathrm{H}}+\sigma_{\mathrm{ul}}^2\bm{\mathrm{I}}_M)^{-1}\bm{\mathrm{h}}_k.
\end{equation}

The optimum combining can only be calculated with channel information of all APs, which can be impractical in a real-world CF-mMIMO system as it requires the BS to combine the channel information from all APs and do matrix inversion. In contrast, a simple MR combining can be calculated locally for each AP by:

\begin{equation} \label{MR}
    \bm{\mathrm{v}}_k = \bm{\mathrm{h}}_k.
\end{equation}
Its downside is the reduced ability to suppress interference for finite array sizes.

\begin{figure}[!t]
    \centering
    \subfloat[L=64]{\includegraphics[width=0.8\linewidth]{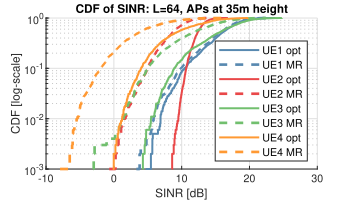}%
    \label{fig:sinr_dist1}}
    \\
    \subfloat[L=256]{\includegraphics[width=0.8\linewidth]{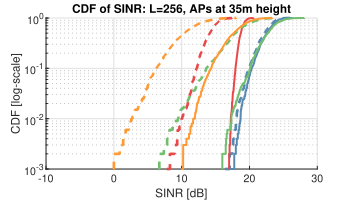}%
    \label{fig:sinr_dist2}}
    
    \caption{SINR with varying number of APs for two combining methods: the optimum combining (Eq. (\ref{SNR})) and the MR combining (Eq. (\ref{MR})) }
    \label{fig:sinr_dist}
\end{figure}

Fig. \ref{fig:sinr_dist} shows the comparison of the SINR values when 64 and 256 APs at 35m height are selected among the trajectory to serve four UEs simultaneously under the two combinings. The first observation is that for UE1, the loss from interference is minimal, with optimum and MR combining curves close to one another. This is because UE1 has the highest channel gains toward APs located at the south of the trajectory, while UE2/3/4 all have low channel gains toward the APs at the south, as shown in Fig. \ref{meas_env}. Thus, the APs that have high weights for reception of UE1 inherently receive little interference power from UE2/3/4. 

The second observation is that UE2/3/4 all experience some loss from interference, shown by the distinct difference between optimum and MR combining. The APs toward the east/north/west of the campus can all receive moderate to high power from UE2/3/4, which act as interference for one another. While UE2 (with $L=64$) has the least outage probability (for bottom 10\% UEs) for the optimum combining due to good channel gains from many APs when the interference can be cancelled, it has worse performance than UE1 for MR combining due to the interference from UE3 and UE4. %Third, we find that UE2 has the steepest curves, as it has more uniform quality of channels to APs all around the campus, while other UEs favor only certain regions of the campus. 
Finally we observe that UE4 generally has relatively smaller channel gains than UE3, due to being under the protruding roof, except for the northeast corner of the 70m trajectory where the AP is at LOS only from UE4. 

%Fig. \ref{fig:sinr_dist2} shows the same SNR distributions, but different SINR distributions achieved through simpler MR combining as discussed above. 
In contrast to the optimum combining, the simpler MR combining generally shows much worse performance, and the gap is expected to increase with the increasing number of UEs. MR combining, despite its simplicity, might not be sufficient for the multi-user cases for all UE2/3/4 even when the number of APs is 256 as shown on Fig. \ref{fig:sinr_dist2}, so combining with higher performance than MR, yet computationally more efficient than optimum combining must be developed. This is remarkable insofar as theory for concentrated massive MIMO says that for the limit of infinitely large arrays, MR becomes optimum. %Clearly, in our case, 64 elements is not large enough. 

%\section{Open Source Channel Data}
%To encourage researchers to use real-world channel data in verifying their CF-mMIMO analysis (or any other topics which can make use of such data), we post our data to \cite{wides}. The channel data is multi-dimensional, containing complex channel values between UE and thousands of AP positions at 1841 frequency points and 128 antenna ports. Any questions regarding the data can be emailed to the corresponding author of this paper.

\section{Conclusion and Future Work}
In this paper, we describe a novel channel sounding method for CF-mMIMO systems: using a drone to measure channels at any desired AP location. Such method provides a large dataset in a short period of time and costs little in comparison to the full setups with many antennas distributed across a wide area. We demonstrate a sample measurement campaign and analyze parameters such as channel gain, SNR, and SINR, to provide some insights on realizing CF-mMIMO systems, such as height of APs, number of APs, and combinings APs may use. We also distribute the real-world channel data open source for various other wireless system analysis. In the future, we will extend the measurement to a larger number of users and environments, in order to provide more statistical evaluations of the CF-mMIMO systems based on real-world data.

\bibliography{IEEEabrv,reference.bib} 
\bibliographystyle{IEEEtran}
\end{document}